\documentclass{article}
\usepackage{amsmath}
\usepackage{amssymb}
\usepackage{amsfonts}
\usepackage{tikz}
\usepackage{latexsym,longtable}
\usepackage[colorlinks,linkcolor=blue,anchorcolor=blue,citecolor=green,CJKbookmarks=True]{hyperref}
\usepackage{tikz}

\newtheorem{theorem}{Theorem}[section]

\newtheorem{lemma}[theorem]{Lemma}

\def\whitebox{{\hbox{\hskip 1pt
 \vrule height 6pt depth 1.5pt
 \lower 1.5pt\vbox to 7.5pt{\hrule width
    3.2pt\vfill\hrule width 3.2pt}%
 \vrule height 6pt depth 1.5pt
 \hskip 1pt } }}
\def\qed{\ifhmode\allowbreak\else\nobreak\fi\hfill\quad\nobreak
     \whitebox\medbreak}
\newcommand{\proof}{\noindent{\it Proof:}\ }

\newcommand{\ignore}[1]{}
\begin{document}

\title{\bf Quantum Latin squares with all possible cardinalities\thanks{Research supported by  NSFC Grant
12271390. (Corresponding author: Lijun Ji)}}

\author{
{\small Ying Zhang, Xin Wang, Lijun Ji}\\
{\small Department of Mathematics}, {\small Soochow University},
{\small Suzhou  215006, China}\\
\texttt{yingzhangzy2025@163.com; xinw@suda.edu.cn; jilijun@suda.edu.cn}\\
}

\date{}
\maketitle
\begin{abstract}
\noindent\noindent
A quantum Latin square of order $n$ (denoted as QLS$(n)$) is an $n\times n$ array whose entries are unit column vectors from the $n$-dimensional Hilbert space $\mathcal{H}_n$, such that each row and column forms an orthonormal basis. Two unit vectors $|u\rangle, |v\rangle\in  \mathcal{H}_n$ are regarded as identical if there exists a real number $\theta$ such that
$|u\rangle=e^{i\theta}|v\rangle$; otherwise, they are considered distinct.
The cardinality $c$ of a QLS$(n)$ is the number of distinct vectors in the
array. In this paper, we use sub-QLS$(4)$s to prove that for any integer $m\geq 2$ and any integer $c\in [4m,16m^2]\setminus \{4m+1\}$, there is a QLS$(4m)$ with cardinality $c$.

 \noindent {\bf Keywords}: \ Latin square, quantum Latin square, cardinality

\end{abstract}

\section{Introduction}

A quantum Latin square of order $n$, denoted as QLS$(n)$, is an $n\times n$ array whose entries are unit column vectors from the $n$-dimensional Hilbert space $\mathcal{H}_n$, such that each row and column forms an orthonormal basis. For any classical Latin square of order $n$ over $\{0,\ldots,n-1\}$, it is possible to obtain a quantum Latin square of order $n$ by interpreting a number $i\in \{0,1,\ldots,n-1\}$ as a column vector $|i\rangle$ in $\mathcal{H}_n$. Specifically,
$|i\rangle$ is a unit vector with its $(i+1)$th component equal to 1, and $\{|0\rangle, |1\rangle, \ldots, |n-1\rangle\}$ is an
orthonormal basis for $\mathcal{H}_n$, referred to as computational basis.
In 2016, Musto and Vicary introduced
quantum Latin squares as a quantum-theoretic generalization of classical Latin squares,
showing their utility in constructing unitary error bases (UEBs) \cite{MV2016}.

In quantum theory \cite{NC2010}, for any two unit vectors $|u\rangle, |v\rangle\in  \mathcal{H}_n$, if there exists a real number $\theta$ such that
$$|u\rangle=e^{i\theta}|v\rangle,$$
then $|u\rangle$ and $|v\rangle$ are considered to represent the same quantum state and are thus regarded
as identical; otherwise, they are considered distinct. 
The cardinality $c$ of a QLS$(n)$ is the number of distinct vectors in the
array. Since a LS$(n)$ always exists, there is a QLS$(n)$ with cardinality $n$. Clearly the cardinality $c$ of a QLS$(n)$ satisfies that $n\leq c\leq  n^2$. A QLS$(n)$ is called apparently quantum if $c = n$, or genuinely quantum if $n<c\leq n^2$. Genuinely quantum Latin squares of order 2 or 3 cannot exist and the possible cardinalities of a quantum Latin squares of order 4 are 4, 6, 8, and 16 \cite{PWRBZ2021}.

In 2021, Nechita and Pillet \cite{NP2021} proposed the concept of quantum Sudoku, a special
case of quantum Latin squares. An $n^2\times n^2$ quantum Sudoku is a QLS$(n^2)$ partitioned
into $n$ disjoint $n\times n$ blocks, with the additional requirement that each block forms an
orthonormal basis of $\mathcal{H}_{n^2}$. In the same year, Paczos et al. \cite{PWRBZ2021} introduced the cardinality
measure for quantum Latin squares, proving the existence of $n^2\times n^2$ quantum Sudoku with maximal cardinality for any positive integer $n$.  Recently, 
Zhang and Cao has almost determined the existence of QLS$(n)$ with maximal cardinality $n^2$ with a few possible exceptions and two definite exceptions $n=2,3$ \cite{ZCPreprint}. For further reading on quantum
theory and quantum Latin squares, the reader is referred to \cite{CLV2024,GRDZ2018,LNV2023,LW2018,RRKL2023,ZTFZ2023,ZTZF2022}.
In this paper, we construct QLS$(4m)$s with all possible cardinalities. 

\begin{theorem}\label{MainResult}
For any integer $m\geq 2$ and any integer $c\in [4m,16m^2]\setminus \{4m+1\}$, there is a QLS$(4m)$ over $\mathcal{H}_m\otimes\mathcal{H}_2^{\otimes 2}$ with cardinality $c$.
\end{theorem}

The remainder of this note is arranged as follows. In Section 2, we present some preliminary results about QLS$(4)$s. In Section 3, we use sub-QLS$(4)$s to QLS$(4m)$s with all possible cardinalities. Finally, we summarize our paper in Section 4.

\section{Preliminaries}

For $i,j\in \{0,1\}$, let $|ij\rangle$ be the tensor product of $|i\rangle$ and $|j\rangle$ from $\mathcal{H}_2$. For any  $a\in \mathbb{R}$, define 
\begin{equation}\label{A_aB_a}
\begin{array}{l}
A_a=\frac{1}{\sqrt{1+a^2}}\begin{pmatrix} |00\rangle+a|01\rangle & -a|00\rangle+|01\rangle\\
	-a|00\rangle+|01\rangle & |00\rangle+a|01\rangle\\
\end{pmatrix},\smallskip \\ B_a=\frac{1}{\sqrt{1+a^2}}\begin{pmatrix} |10\rangle+a|11\rangle & -a|10\rangle+|11\rangle\\
	-a|10\rangle+|11\rangle & |10\rangle+a|11\rangle\\
\end{pmatrix}.
\end{array}	
\end{equation}
Clearly, each $A_a$ is a QLS$(2)$ over the subspace $\mathcal{L}(|00\rangle,|01\rangle)=\{c_1|00\rangle+c_2|01\rangle\colon c_1,c_2\in \mathbb{C}\}$, while $B_a$ is a QLS$(2)$ over $\mathcal{L}(|10\rangle,|11\rangle)=\{c_1|10\rangle+c_2|11\rangle\colon c_1,c_2\in \mathbb{C}\}$. Hence, the $4\times 4$ matrix of the form $\begin{pmatrix}
	A_c & B_x\\
	B_y & A_d\\
\end{pmatrix}$ with arbitrary $c,d,x,y\in \mathbb{R}$ is a QLS$(4)$ in $\mathcal{H}_2^{\otimes 2}$.

\begin{lemma}\label{distinctQLS(4)}
Let $H_0=\begin{pmatrix}
	A_0 & B_0\\
	B_0 & A_0\\
\end{pmatrix}$ and $H_1=\begin{pmatrix}
A_0 & B_0\\
B_1 & A_1\\
\end{pmatrix}$, where $A_a$ and $B_a$ are defined in Equation $(\ref{A_aB_a})$. Then
there is a QLS$(4)$ in $\mathcal{H}_2^{\otimes 2}$ with exactly $\ell$ distinct elements neither from $H_0$ nor from $H_1$ for $\ell\in \{2,3,4,5,6,7,8\}$.
\end{lemma}

\proof  
 Define 
 \begin{equation}\label{J1}
 	J_1=\begin{pmatrix}
 		1 & 0 & 0 & 0 \\
 		0 & \frac{1}{3} & -\frac{2}{3} & \frac{2}{3} \smallskip \\
 		0 & \frac{2}{3} & -\frac{1}{3} & -\frac{2}{3} \smallskip \\
 		0 & \frac{2}{3} & \frac{2}{3} & \frac{1}{3} 
 	\end{pmatrix}.
 \end{equation}
Clearly, $J_1$ is an orthonormal matrix. It follows that the vectors $\alpha_1,\alpha_2,\alpha_3,\alpha_4$ defined by $(\alpha_1,\alpha_2,\alpha_3,\alpha_4)=(|00\rangle, |01\rangle, |10\rangle,|11\rangle)J_1$ also form an orthonormal basis for $\mathcal{H}_2^{\otimes 2}$. Note that $\alpha_1=|00\rangle$.
 
 For $a\in \mathbb{R}$, define 
 $$C_a=\frac{1}{\sqrt{1+a^2}}\begin{pmatrix} \alpha_1+a\alpha_2 & -a\alpha_1+\alpha_2\\
 	-a\alpha_1+\alpha_2 & \alpha_1+a\alpha_2\\
 \end{pmatrix}, \ D_a=\frac{1}{\sqrt{1+a^2}}\begin{pmatrix} \alpha_3+a\alpha_4 & -a\alpha_3+\alpha_4\\
 	-a\alpha_3+\alpha_4 & \alpha_3+a\alpha_4\\
 \end{pmatrix}.$$
 Clearly, each $C_a$ is a QLS$(2)$ over $\mathcal{L}(\alpha_1,\alpha_2)=\{c_1\alpha_1+c_2\alpha_2\colon c_1,c_2\in \mathbb{C}\}$, while $D_a$ is also a QLS$(2)$ over $\mathcal{L}(\alpha_3,\alpha_4)=\{c_1\alpha_3+c_2\alpha_4\colon c_1,c_2\in \mathbb{C}\}$. Hence, the $4\times 4$ matrix of the form $\begin{pmatrix}
 	C_a & D_x\\
 	D_y & C_b\\
 \end{pmatrix}$ with arbitrary $a,b,x,y\in \mathbb{R}$ is a QLS$(4)$ in $\mathcal{H}_2^{\otimes 2}$. 

According to  $\alpha_2=\frac{1}{3}|01\rangle+\frac{2}{3}|10\rangle+\frac{2}{3}|11\rangle$, we have $\mathcal{L}(|00\rangle,|01\rangle)\cap \mathcal{L}(\alpha_1,\alpha_2)=\mathcal{L}(|00\rangle)$ and $\mathcal{L}(|10\rangle,|11\rangle)\cap \mathcal{L}(\alpha_1,\alpha_2)$ contains only zero vector. It follows that all elements in $C_a$ differ from those in $B_x$ for any $a,x\in \mathbb{R}$ and that all elements in $C_a$ differ from those in $A_x$ for any $a,x\in \mathbb{R}$ with $(a,x)\neq (0,0)$. Also, $A_0$ and $C_0$ has exactly one common element $|00\rangle$. 

According to  $\alpha_3=-\frac{2}{3}|01\rangle-\frac{1}{3}|10\rangle+\frac{2}{3}|11\rangle$ and $\alpha_4=\frac{2}{3}|01\rangle-\frac{2}{3}|10\rangle+\frac{1}{3}|11\rangle$, we have $\mathcal{L}(|00\rangle,|01\rangle)\cap \mathcal{L}(\alpha_3,\alpha_4)=\{\bar{0}\}$ and $\mathcal{L}(|10\rangle,|11\rangle)\cap \mathcal{L}(\alpha_3,\alpha_4)=\mathcal{L}(|10\rangle-|11\rangle)$. It follows that all elements in $D_a$ differ from those in $A_x$ for any $a,x\in \mathbb{R}$ and that all elements in $D_a$ differ from those in $B_x$ for any $a,x\in \mathbb{R}$ with $(a,x)\neq (1,-1)$. Also, $B_{-1}$ and $D_1$ has exactly one common element $\frac{1}{\sqrt{2}}(|10\rangle-|11\rangle)$. 

By the definition of $A_a$, it is easy to see that the elements in $A_a$ differ from those in $A_x$ for any distinct $a,x\in \mathbb{R}$. Similarly, the elements in $B_a$ differ from those in $B_x$, the elements in $C_a$ differ from those in $C_x$, and the elements in $D_a$ differ from those in $D_x$ for any distinct $a,x\in \mathbb{R}$.  
 
For $\ell\in \{2,3,\ldots,8\}$, we define a QLS$(4)$ $H_{\ell}$ in the following table with exactly $\ell$ distinct elements neither from $H_0$ nor from $H_1$. 
 
 \begin{center}
 	
 	\begin{tabular}{|c|c|c|c|c|c|c|}
 		\hline
 		$H_2$ & $H_3$ & $H_4$ & $H_5$ & $H_6$ & $H_7$ & $H_8$ \\ \hline 
 		$\begin{pmatrix}
 			A_0 & B_0\\
 			B_0 & A_2\\
 		\end{pmatrix}$ & $\begin{pmatrix}
 			C_0 & D_0\\
 			D_0 & C_0\\
 		\end{pmatrix}$ & $\begin{pmatrix}
 			A_0 & B_0\\
 			B_2 & A_2\\
 		\end{pmatrix}$ & $\begin{pmatrix}
 			C_0 & D_0\\
 			D_0 & C_1\\
 		\end{pmatrix}$  & $\begin{pmatrix}
 			A_0 & B_2\\
 			B_3 & A_2\\
 		\end{pmatrix}$ & $\begin{pmatrix}
 			C_0 & D_3\\
 			D_4 & C_1\\
 		\end{pmatrix}$ &  $\begin{pmatrix}
 			A_2 & B_2\\
 			B_3 & A_3\\
 		\end{pmatrix}$ \\ \hline
 	\end{tabular}
 \end{center} 

For $H_5$, the numbers of distinct elements in $C_0, C_1$ and $D_0$ differing from those in $H_0$ are $1, 2$ and $2$, respectively, thereby it has five distinct elements totally differing from those in $H_0$. Similarly,  it has five distinct elements totally differing from those in $H_1$. The same discuss counts the number of distinct elements in $H_{\ell}$ differing from those in $H_0$ or in $H_1$ for $\ell \in\{2,3,4,6,7,8\}$. This completes the proof. \qed

An $m\times n$ row-quantum Latin rectangle is an array with $m$ rows and $n$ columns whose entries are unit vectors from the $n$-dimensional Hilbert space $\mathcal{H}_n$, and such that
the elements in each row are mutually orthogonal.

Let $U$ and $V$ denote an $m\times n$ row-quantum Latin rectangle with cardinality $c_1$ in $\mathcal{H}_n$ and
an $n\times m$ row-quantum Latin rectangle with cardinality $c_2$ in $\mathcal{H}_m$, respectively. Let $|u_{i,j}\rangle$ and $|v_{k,l}\rangle$
represent the elements in the $i$th row and $j$th column of $U$, and the $k$th row and $l$th column
of $V$, respectively. Let $W$ be a matrix of order $mn$, which is divided into $mn$ blocks of size $n\times m$. Denote by
$|w_{i,j;k,l}\rangle$ the element located in the $k$th row and $l$th column of the $(i,j)$th block in the
matrix $W$ and define the tensor product
\begin{equation}\label{matrixtensorproduct}
	|w_{i,j;k,l}\rangle =|u_{i,j+k}\rangle \otimes |v_{j,i+l}\rangle
\end{equation}
where the addition $j + k$ is modulo $n$ and $i + l$ is modulo $m$. 

\begin{lemma}{\rm \cite[Theorem 3.4]{ZCPreprint}}\label{ProductConstruction}
	Let $U$ and $V$ denote an $m\times n$ row-quantum Latin rectangle with cardinality $c_1$ in $\mathcal{H}_n$ and
	an $n\times m$ row-quantum Latin rectangle with cardinality $c_2$ in $\mathcal{H}_m$, respectively. Then the $mn\times mn$ array $W$ defined by Equation $(\ref{matrixtensorproduct})$ is a QLS$(mn)$ with cardinality $c_1c_2$.
\end{lemma}
 
For $a,b\in \mathbb{R}$ with $a\neq b$, define 
\begin{equation}\label{Vab}
	V_{a,b}=
	\begin{pmatrix}
		\displaystyle \frac{1}{\sqrt{1+a^2}}|0\rangle + \displaystyle \frac{a}{\sqrt{1+a^2}}|1\rangle & \displaystyle -\frac{a}{\sqrt{1+a^2}}|0\rangle +\frac{1}{\sqrt{1+a^2}}|1\rangle \smallskip \\
		\displaystyle \frac{1}{\sqrt{1+b^2}}|0\rangle + \displaystyle \frac{b}{\sqrt{1+b^2}}|1\rangle & \displaystyle -\frac{b}{\sqrt{1+b^2}}|0\rangle +\frac{1}{\sqrt{1+b^2}}|1\rangle 
	\end{pmatrix}.
\end{equation}
Then $V_{a,b}$ is a row-quantum Latin rectangle in $\mathcal{H}_2$ with cardinality $4$. By applying Lamma \ref{ProductConstruction} with $V_{0,1}$ and $V_{a,b}$, a $4\times 4$ matrix $W_{a,b}$ defined by Equation $(\ref{matrixtensorproduct})$ is a QLS$(4)$ with maximal cardinality 16, which is displayed below for use later,
\begin{equation}\label{Wab}
	W_{a,b}=\begin{pmatrix}
		(|00\rangle, |01\rangle, |10\rangle, |11\rangle)Y_{1,a,b}\\
		(|00\rangle, |01\rangle, |10\rangle, |11\rangle)Y_{2,a,b}\\
		(|00\rangle, |01\rangle, |10\rangle, |11\rangle)Y_{3,a,b}\\
		(|00\rangle, |01\rangle, |10\rangle, |11\rangle)Y_{4,a,b}\\	
	\end{pmatrix},	
\end{equation}
where $Y_{i,a,b}$ are as follows:
\begin{equation}\label{Yiab}
	\begin{array}{l}
		Y_{1,a,b}=\begin{pmatrix}
			\frac{1}{\sqrt{1+a^2}} & -\frac{a}{\sqrt{1+a^2}} & 0 & 0\smallskip \\
			\frac{a}{\sqrt{1+a^2}} & \frac{1}{\sqrt{1+a^2}} & 0 & 0\\				
			0 & 0 & \frac{1}{\sqrt{1+b^2}} & -\frac{b}{\sqrt{1+b^2}}\smallskip \\
			0 & 0 & \frac{b}{\sqrt{1+b^2}} & \frac{1}{\sqrt{1+b^2}}\\
		\end{pmatrix}, \smallskip \\
		Y_{2,a,b}=\begin{pmatrix}
			0 & 0 & \frac{1}{\sqrt{1+b^2}} & -\frac{b}{\sqrt{1+b^2}}\smallskip \\
			0 & 0 & \frac{b}{\sqrt{1+b^2}} & \frac{1}{\sqrt{1+b^2}}\\
			\frac{1}{\sqrt{1+a^2}} & -\frac{a}{\sqrt{1+a^2}} & 0 & 0\smallskip \\
			\frac{a}{\sqrt{1+a^2}} & \frac{1}{\sqrt{1+a^2}} & 0 & 0\\				
		\end{pmatrix}, \smallskip \\
		Y_{3,a,b}=\begin{pmatrix}
			-\frac{a}{\sqrt{2+2a^2}} & \frac{1}{\sqrt{2+2a^2}} & \frac{b}{\sqrt{2+2b^2}} & -\frac{1}{\sqrt{2+2b^2}}\smallskip \\
			\frac{1}{\sqrt{2+2a^2}} & \frac{a}{\sqrt{2+2a^2}} & -\frac{1}{\sqrt{2+2b^2}} & -\frac{b}{\sqrt{2+2b^2}}\smallskip \\				
			-\frac{a}{\sqrt{2+2a^2}} & 	\frac{1}{\sqrt{2+2a^2}} & -\frac{b}{\sqrt{2+2b^2}} & \frac{1}{\sqrt{2+2b^2}}\smallskip \\
			\frac{1}{\sqrt{2+2a^2}} & 	\frac{a}{\sqrt{2+2a^2}} & \frac{1}{\sqrt{2+2b^2}} & \frac{b}{\sqrt{2+2b^2}}\\
		\end{pmatrix}, \smallskip \\
		Y_{4,a,b}=\begin{pmatrix}
			\frac{a}{\sqrt{2+2a^2}} & -\frac{1}{\sqrt{2+2a^2}} & -\frac{b}{\sqrt{2+2b^2}} & \frac{1}{\sqrt{2+2b^2}}\smallskip \\
			-\frac{1}{\sqrt{2+2a^2}} & -\frac{a}{\sqrt{2+2a^2}} & \frac{1}{\sqrt{2+2b^2}} & \frac{b}{\sqrt{2+2b^2}}\smallskip \\				
			-\frac{a}{\sqrt{2+2a^2}} & 	\frac{1}{\sqrt{2+2a^2}} & -\frac{b}{\sqrt{2+2b^2}} & \frac{1}{\sqrt{2+2b^2}}\smallskip \\
			\frac{1}{\sqrt{2+2a^2}} & 	\frac{a}{\sqrt{2+2a^2}} & \frac{1}{\sqrt{2+2b^2}} & \frac{b}{\sqrt{2+2b^2}}\\
		\end{pmatrix}.
	\end{array}	
\end{equation}
Note that each $Y_{i,a,b}$ is an orthonormal matrix for $1\leq i\leq 4$ and all of the $j$th column vectors of $Y_{i,a,b}$, $1\leq i\leq 4$, form an orthonormal basis in $\mathbb{R}^4$ for $1\leq j\leq 4$.

\begin{lemma}\label{infiniteQSL(4)Cardinality16}
	Let $W_{a,b}$ be defined in Equation $(\ref{Wab})$. Then $W_{2k-1,2k}$, $k\in \mathbb{N}$, are QLS$(4)$s with maximal cardinality $16$ in $\mathcal{H}_2^{\otimes 2}$ such that all $32$ elements of every two QLS$(4)$s are distinct.
\end{lemma}

\proof As described above, each $W_{2k-1,2k}$ is a  QLS$(4)$ with maximal cardinality $16$ in $\mathcal{H}_2^{\otimes 2}$. For $k,t\in \mathbb{N}$ with $k\neq t$, since $k-t\neq 0$, $1+kt\neq 0$ and $1-kt\neq 0$, all $32$ elements of $W_{2k-1,2k}$ and $W_{2t-1,2t}$ are distinct. \qed

\section{QLS$(4m)$ with all possible cardinalities}

In this section we use sub-QLS$(4)$s to construct QLS$(4m)$s with all possible cardinalities.

\begin{lemma}\label{n+1}
	For any integer $n\geq 2$, there does not exist a QLS$(n)$ with cardinality $n+1$.	
\end{lemma}

\proof Assume on the contrary that $A$ is QLS$(n)$ with cardinality $n+1$ in $\mathcal{H}_n$. Let the first row consist of vectors $|0\rangle, |1\rangle, \ldots, |n-1\rangle$. There is a row with exactly one element $\alpha$ distinct from the elements in the first row. Suppose that this row contains vectors $|0\rangle, |1\rangle, \ldots, |n-2\rangle$. Other cases can be proved similarly. Since $\alpha$ is orthogonal to $|0\rangle, |1\rangle, \ldots, |n-2\rangle$, we have  $\alpha=|n-1\rangle$, a contradiction. 
\qed

\begin{lemma}\label{QLS(8)}
For any integer $c\in [8,64]\setminus \{9\}$, there is a QLS$(8)$ in $\mathcal{H}_2^{\otimes 3}$ with cardinality $c$.
\end{lemma}

\proof  By Lemma \ref{n+1}, $c\neq 9$. Let $H_0$ and $H_1$ be defined as in Lemma \ref{distinctQLS(4)} in $\mathcal{H}_2^{\otimes 2}$. By Lemma \ref{distinctQLS(4)}, there is a QLS$(4)$ $H_{\ell}$ with exactly $\ell$ distinct elements neither from $H_0$ nor from $H_1$ for $\ell\in \{2,3,\ldots,8\}$. By comparing the coordinate vectors of elements in $W_{a,b}$ and $H_{\ell}$ for $\ell\in \{0,1,\ldots,8\}$, it is easy to check that there are two QSL$(4)$s, say $W_{5,6}$ and $W_{7,8}$ in Lemma \ref{infiniteQSL(4)Cardinality16}, in $\mathcal{H}_2^{\otimes 2}$ with cardinality 16 and these 32 distinct elements differ from $H_{\ell}$ for any $\ell\in \{0,1,\ldots,8\}$. 

Define QLS$(8)$s in the following table where $\ell\in \{0,2,3,\ldots,8\}$. Simple counting gives the corresponding cardinalities.
\begin{center}
	\begin{tabular}{|c|c|c|c|}
		\hline
		QLS$(8)$ & $\begin{pmatrix}
			|0\rangle \otimes H_0 & |1\rangle \otimes H_0\\
			|1\rangle \otimes H_0 & |0\rangle \otimes H_{\ell}\\
		\end{pmatrix}$ & $\begin{pmatrix}
			|0\rangle \otimes H_0 & |1\rangle \otimes H_8\\
			|1\rangle \otimes H_8 & |0\rangle \otimes H_{\ell}\\
		\end{pmatrix}$ & $\begin{pmatrix}
			|0\rangle \otimes H_0 & |1\rangle \otimes H_{0}\\
			|1\rangle \otimes H_{8} & |0\rangle \otimes H_{\ell}\\
		\end{pmatrix}$ \\ \hline 
		Cardinality & $8+\ell$ & $12+\ell$ & $16+\ell$ \\ \hline \hline
		QLS$(8)$ & $\begin{pmatrix}
			|0\rangle \otimes H_{0} & |1\rangle \otimes W_{5,6}\\
			|1\rangle \otimes W_{5,6} & |0\rangle \otimes H_{\ell}\\
		\end{pmatrix}$ & $\begin{pmatrix}
			|0\rangle \otimes H_0 & |1\rangle \otimes H_0\\
			|1\rangle \otimes W_{5,6} & |0\rangle \otimes H_{\ell}\\
		\end{pmatrix}$ & $\begin{pmatrix}
			|0\rangle \otimes H_0 & |1\rangle \otimes H_{8}\\
			|1\rangle \otimes W_{5,6} & |0\rangle \otimes H_{\ell}\\
		\end{pmatrix}$ \\ \hline 
		Cardinality & $20+\ell$ & $24+\ell$ & $28+\ell$ \\ \hline \hline
		QLS$(8)$ & $\begin{pmatrix}
			|0\rangle \otimes H_{1} & |1\rangle \otimes H_{8}\\
			|1\rangle \otimes W_{5,6} & |0\rangle \otimes H_{\ell}\\
		\end{pmatrix}$ & $\begin{pmatrix}
			|0\rangle \otimes H_0 & |1\rangle \otimes W_{5,6}\\
			|1\rangle \otimes W_{7,8} & |0\rangle \otimes H_{\ell}\\
		\end{pmatrix}$ & $\begin{pmatrix}
			|0\rangle \otimes H_1 & |1\rangle \otimes W_{5,6}\\
			|1\rangle \otimes W_{7,8} & |0\rangle \otimes H_{\ell}\\
		\end{pmatrix}$ \\ \hline 
		Cardinality & $32+\ell$ & $36+\ell$ & $40+\ell$ \\ \hline
	\end{tabular}
\end{center}
Then for $c\in [8,48]\setminus \{9\}$, there is a QLS$(8)$ in $\mathcal{H}_2^{\otimes 3}$ with cardinality $c$. 

By applying Lemma \ref{ProductConstruction} with $V_{0,4/3}$ and $V_{0,12/5}$ defined in Equation (\ref{Vab}), i.e.,
\begin{equation*}
	V_{0,4/3}=
	\begin{pmatrix}
		\displaystyle |0\rangle & |1\rangle \smallskip\\ \displaystyle \frac{3}{5}|0\rangle +\frac{4}{5}|1\rangle &
		\displaystyle -\frac{4}{5}|0\rangle + \displaystyle \frac{3}{5}|1\rangle
	\end{pmatrix},\ 
	V_{0,12/5}=
	\begin{pmatrix}
		\displaystyle |0\rangle & |1\rangle \smallskip\\ \displaystyle \frac{5}{13}|0\rangle +\frac{12}{13}|1\rangle &
		\displaystyle -\frac{12}{13}|0\rangle + \displaystyle \frac{5}{13}|1\rangle
	\end{pmatrix},
\end{equation*}
 a $4\times 4$ matrix $W_{0}$ defined by Equation $(\ref{matrixtensorproduct})$ is a QLS$(4)$ with maximal cardinality 16, which is listed below for use later,
\begin{equation}\label{W0}
	W_{0}=\begin{pmatrix}
		(|00\rangle, |01\rangle, |10\rangle, |11\rangle)X_{1}\\
		(|00\rangle, |01\rangle, |10\rangle, |11\rangle)X_{2}\\
		(|00\rangle, |01\rangle, |10\rangle, |11\rangle)X_{3}\\
		(|00\rangle, |01\rangle, |10\rangle, |11\rangle)X_{4}\\	
	\end{pmatrix}	
\end{equation}
where $X_{i}$ are as follows:
\begin{equation*}
	X_1 = \begin{pmatrix}
		0 & 0 & -\frac{12}{13} & \frac{5}{13} \smallskip \\
		0 & 0 & \frac{5}{13} & \frac{12}{13} \smallskip \\
		0 & 1 & 0 & 0 \\
		1 & 0 & 0 & 0 
	\end{pmatrix}, \
	X_2 = \begin{pmatrix} 
		-\frac{4}{5} & 0 & \frac{3}{13} & -\frac{36}{65} \smallskip \\
		0 & -\frac{4}{5} & \frac{36}{65} & \frac{3}{13} \smallskip \\
		\frac{3}{5} & 0 & \frac{4}{13} & -\frac{48}{65} \smallskip \\
		0 & \frac{3}{5} & \frac{48}{65} & \frac{4}{13} 
	\end{pmatrix},
\end{equation*}
\begin{equation*}
	X_3 = \begin{pmatrix} 
		0 & 1 & 0 & 0 \\
		1 & 0 & 0 & 0 \\
		0 & 0 & -\frac{12}{13} & \frac{5}{13} \smallskip \\
		0 & 0 & \frac{5}{13} & \frac{12}{13}
	\end{pmatrix},\
	X_4 = \begin{pmatrix} 
		\frac{3}{5} & 0 & -\frac{4}{13} & \frac{48}{65} \smallskip \\
		0 & \frac{3}{5} & -\frac{48}{65} & -\frac{4}{13} \smallskip \\
		\frac{4}{5} &0 & \frac{3}{13} & -\frac{36}{65} \smallskip \\
		0 & \frac{4}{5} & \frac{36}{65} & \frac{3}{13}
	\end{pmatrix}.
\end{equation*}
Note that each $X_{i}$ is an orthonormal matrix for $1\leq i\leq 4$ and all of the $j$th column vectors of $X_{i}$, $1\leq i\leq 4$, form an orthonormal basis in $\mathbb{R}^4$ for $1\leq j\leq 4$.

Define 

\begin{equation}\label{J2}
	J_2 = \begin{pmatrix}
		1 & 0 & 0 & 0 \\
		0 & 1 & 0 & 0 \\
		0 & 0 & \frac{3}{5} & -\frac{4}{5} \smallskip \\
		0 & 0 & \frac{4}{5} & \frac{3}{5}
	\end{pmatrix},\ 
	J_3=\begin{pmatrix}
		1 & 0 & 0 & 0\\
		0 & \frac{3}{5} & -\frac{4}{5} & 0 \smallskip \\
		0 & \frac{4}{5} & \frac{3}{5} & 0\\
		0 & 0 & 0 & 1\\
	\end{pmatrix},\
	J_4=\begin{pmatrix}
		1 & 0 & 0 & 0\\
		0 & 1 & 0 & 0\\
		0 & 0 & -\frac{4}{5} & \frac{3}{5} \smallskip \\
		0 & 0 & \frac{3}{5} & \frac{4}{5}\\
	\end{pmatrix}.
\end{equation}

For $k\in \{1,2,3,4\}$, define
\begin{equation}\label{W0}
	W_{k}=\begin{pmatrix}
		(|00\rangle, |01\rangle, |10\rangle, |11\rangle)X_1J_kX_1^{-1}X_1\\
		(|00\rangle, |01\rangle, |10\rangle, |11\rangle)X_1J_kX_1^{-1}X_2\\
		(|00\rangle, |01\rangle, |10\rangle, |11\rangle)X_1J_kX_1^{-1}X_3\\
		(|00\rangle, |01\rangle, |10\rangle, |11\rangle)X_1J_kX_1^{-1}X_4\\	
	\end{pmatrix}.	
\end{equation}
Since all $X_1, J_k$ are orthonormal matrices, the vectors $(|00\rangle, |01\rangle, |10\rangle,|11\rangle)X_1J_kX_1^{-1}$ also form an orthonormal basis for $\mathcal{H}_2^{\otimes 2}$. Since the column vectors of $X_1, X_2, X_3$ and $X_4$ form a QLS$(4)$ with cardinality 16 in $\mathcal{H}_4$, each $W_k$ is a QLS$(4)$ with cardinality 16. Further, simple computation shows that 

\begin{center}
	$ X_1J_1X_1^{-1}X_1=
	\begin{pmatrix}
		0 & -\frac{14}{39} & \frac{22}{39} & \frac{29}{39} \smallskip\\
		0 & \frac{34}{39} & \frac{19}{39} & \frac{2}{39} \smallskip\\
		0 & \frac{1}{3} & -\frac{2}{3} & \frac{2}{3} \smallskip\\
		1 & 0 & 0 & 0\\
	\end{pmatrix}$,
	$X_1J_1X_1^{-1}X_2=
	\begin{pmatrix}
		-\frac{14}{507} & -\frac{1832}{2535} & \frac{851}{2535} & \frac{102}{169} \smallskip\\
		\frac{2198}{2535} & -\frac{476}{2535} & \frac{758}{2535} & -\frac{297}{845} \smallskip\\
		-\frac{97}{195}& -\frac{56}{195} & \frac{98}{195} & -\frac{42}{65} \smallskip\\
		0 & \frac{3}{5} & \frac{48}{65} & \frac{4}{13} \\
	\end{pmatrix}$,
\end{center}
\begin{center}
	$X_1J_1X_1^{-1}X_3=\begin{pmatrix}
		\frac{458}{507} & -\frac{119}{507} & \frac{56}{169} & -\frac{70}{507} \smallskip\\
		\frac{119}{507} & -\frac{218}{507} & -\frac{136}{169} & \frac{170}{507} \smallskip\\
		\frac{14}{39} & \frac{34}{39} & -\frac{4}{13} & \frac{5}{39} \smallskip\\
		0 & 0 & \frac{5}{13} & \frac{12}{13}\\
	\end{pmatrix}$,\
	$X_1J_1X_1^{-1}X_4=\begin{pmatrix}
		-\frac{217}{507} & \frac{458}{845} & -\frac{1718}{2535} & -\frac{128}{507} \smallskip\\
		\frac{1114}{2535} & \frac{119}{845} & \frac{406}{2535} & -\frac{2212}{2535} \smallskip\\
		\frac{154}{195} & \frac{14}{65} & -\frac{89}{195} & \frac{68}{195} \smallskip\\
		0 & \frac{4}{5} & \frac{36}{65} & \frac{3}{13} \\
	\end{pmatrix}$,
\end{center}
\begin{center}
	$X_1J_2X_1^{-1}X_1=
	\begin{pmatrix}
		0 & 0 & -\frac{16}{65} & \frac{63}{65} \smallskip\\
		0 & 0 & \frac{63}{65} & \frac{16}{65} \\
		0 & 1 & 0 & 0 \\
		1 & 0 & 0 & 0 \\
	\end{pmatrix} $,
	$X_1J_2X_1^{-1}X_2=
	\begin{pmatrix}
		-\frac{12}{25} & -\frac{16}{25} & \frac{189}{325} & -\frac{48}{325} \smallskip\\
		\frac{16}{25} & -\frac{12}{25} & \frac{48}{325} & \frac{189}{325} \smallskip\\
		\frac{3}{5} & 0 & \frac{4}{13} & -\frac{48}{65} \smallskip\\
		0 & \frac{3}{5} & \frac{48}{65} & \frac{4}{13} \\
	\end{pmatrix} $,
\end{center}
\begin{center}
	$X_1J_2X_1^{-1}X_3=\begin{pmatrix}
		\frac{4}{5} & \frac{3}{5} & 0 & 0 \smallskip\\
		\frac{3}{5} & -\frac{4}{5} & 0 & 0 \\
		0 & 0 & -\frac{12}{13} & \frac{5}{13} \smallskip\\
		0 & 0 & \frac{5}{13} & \frac{12}{13}\\
	\end{pmatrix}$,
	$X_1J_2X_1^{-1}X_4=\begin{pmatrix}
		\frac{9}{25} & \frac{12}{25} & -\frac{252}{325} & \frac{64}{325} \smallskip\\
		-\frac{12}{25} & \frac{9}{25} & -\frac{64}{325} & -\frac{252}{325} \smallskip\\
		\frac{4}{5} & 0 & \frac{3}{13} & -\frac{36}{65} \smallskip\\
		0 & \frac{4}{5} & \frac{36}{65} & \frac{3}{13}\\
	\end{pmatrix}
	$,
\end{center}
\begin{center}
	$X_1J_3X_1^{-1}X_1=\begin{pmatrix}
		0 & -\frac{48}{65} & -\frac{36}{65} & \frac{5}{13} \smallskip\\
		0 & \frac{4}{13} & \frac{3}{13} & \frac{12}{13} \smallskip\\
		0 & \frac{3}{5} & -\frac{4}{5} & 0 \\
		1 & 0 & 0 & 0 \\
	\end{pmatrix} $,
	$X_1J_3X_1^{-1}X_2=\begin{pmatrix}
		-\frac{164}{169} & -\frac{96}{845} & \frac{3}{845} & \frac{36}{169} \smallskip\\
		\frac{12}{169} & -\frac{636}{845} & \frac{548}{845} & -\frac{15}{169} \smallskip \\
		-\frac{3}{13} & \frac{16}{65} & \frac{12}{65} & -\frac{12}{13} \smallskip\\
		0 & \frac{3}{5} & \frac{48}{65} & \frac{4}{13}  \\
	\end{pmatrix}$,
\end{center}
\begin{center}
	$X_1J_3X_1^{-1}X_3=\begin{pmatrix}
		\frac{24}{169} & \frac{557}{845} & \frac{576}{845} & -\frac{48}{169} \smallskip\\
		\frac{159}{169} & \frac{24}{169} & -\frac{48}{169} & \frac{20}{169} \smallskip\\
		-\frac{4}{13} & \frac{48}{65} & -\frac{36}{65} & \frac{3}{13} \smallskip\\
		0 & 0 & \frac{5}{13} & \frac{12}{13} \\
	\end{pmatrix}$,
	$X_1J_3X_1^{-1}X_4=\begin{pmatrix}
		-\frac{33}{169} & \frac{72}{845} & -\frac{404}{845} & \frac{144}{169} \smallskip\\
		\frac{56}{169} & \frac{477}{845} & -\frac{564}{845} & -\frac{60}{169} \smallskip\\
		\frac{12}{13} & -\frac{12}{65} & \frac{9}{65} & \frac{4}{13} \smallskip\\
		0 & \frac{4}{5} & \frac{36}{65} & \frac{3}{13}\\
	\end{pmatrix}$,
\end{center}
\begin{center}
	$X_1J_4X_1^{-1}X_1=\begin{pmatrix}
		0 & 0 & \frac{63}{65} & -\frac{16}{65} \smallskip\\
		0 & 0 & \frac{16}{65} & \frac{63}{65} \\
		0 & 1 & 0 & 0 \\
		1 & 0 & 0 & 0 \\
	\end{pmatrix}$,
	$X_1J_4X_1^{-1}X_2=\begin{pmatrix}
		\frac{3344}{4225} & -\frac{492}{4225} & -\frac{48}{325} & \frac{189}{325} \smallskip\\
		-\frac{492}{4225} & -\frac{3344}{4225} & \frac{189}{325} & \frac{48}{325} \smallskip\\
		\frac{3}{5} & 0 & \frac{4}{13} & -\frac{48}{65} \smallskip\\
		0 & \frac{3}{5} & \frac{48}{65} & \frac{4}{13}\\
	\end{pmatrix}$, 
\end{center}
\begin{center}
	$X_1J_4X_1^{-1}X_3=\begin{pmatrix}
		\frac{123}{845} & -\frac{836}{845} & 0 & 0 \smallskip\\
		\frac{836}{845} & \frac{123}{845} & 0 & 0 \\
		0 & 0 & -\frac{12}{13} & \frac{5}{13} \smallskip\\
		0 & 0 & \frac{5}{13} & \frac{12}{13} \\
	\end{pmatrix}$,
	$X_1J_4X_1^{-1}X_4=\begin{pmatrix}
		-\frac{2508}{4225} & \frac{369}{4225} & \frac{64}{325} & -\frac{252}{325} \smallskip\\
		\frac{369}{4225} & \frac{2508}{4225} & -\frac{252}{325} & -\frac{64}{325} \smallskip\\
		\frac{4}{5} & 0 & \frac{3}{13} & -\frac{36}{65} \smallskip\\
		0 & \frac{4}{5} & \frac{36}{65} & \frac{3}{13} \\
	\end{pmatrix}$.
\end{center}

It has been checked that 

(P1) $W_0$ and $W_1$ have a common element $|11\rangle$, 

(P2) $W_0$ and $W_2$ have exactly four common elements: $|10\rangle$, $|11\rangle$, $-\frac{12}{13} |10\rangle+\frac{5}{13}|11\rangle$, $\frac{5}{13}|10\rangle+\frac{12}{13}|11\rangle$, 

(P3) $W_0$ and $W_3$ have exactly two common elements: $|11\rangle$,  $\frac{5}{13}|00\rangle+\frac{12}{13}|01\rangle$,

(P4) $W_2$ and $W_4$ have exactly six common elements: $|10\rangle$, $|11\rangle$, $-\frac{63}{65} |00\rangle+\frac{16}{65}|01\rangle$, $-\frac{16}{65}|00\rangle+\frac{63}{65}|01\rangle$,

\qquad  $-\frac{12}{13}|10\rangle+\frac{5}{13}|11\rangle$, $\frac{5}{13}|10\rangle+\frac{12}{13}|11\rangle$.

\noindent Hence, there is a QLS$(4)$ in $\mathcal{H}_2^{\otimes 2}$ with $\ell$ distinct elements not from $W_0$ for $\ell\in \{12,14,15\}$. Since the coefficients of $W_{5,6}$ in Equation (\ref{Wab}) are irrational in orthonormal basis $|00\rangle, |01\rangle, |10\rangle, |11\rangle$, there are 16 distinct elements in $W_{5,6}$ not from $W_0$. 

Define QLS$(4)$ in $\mathcal{H}_2^{\otimes 2}$ as follows:

\begin{center}
	\begin{tabular}{|c|c|c|c|c|}
		\hline
		QLS$(4)$ &
		$\begin{pmatrix}
			B_0 & A_0\\
			A_{1} & B_0\\
		\end{pmatrix}$ & $\begin{pmatrix}
			B_0 & A_0\\
			A_{1} & B_{1}\\
		\end{pmatrix}$ & $\begin{pmatrix}
			B_{1} & A_{2}\\
			A_{1} & B_{0}\\
		\end{pmatrix}$ & $\begin{pmatrix}
			B_{1} & A_{1}\\
			A_{2} & B_{2}\\
		\end{pmatrix}$ \\ \hline
		$\ell$ & 2 & 4 & 6 & 8\\ \hline
	\end{tabular}
\end{center}

\noindent where $A_{x}$ and $B_{x}$ are defined in Equation (\ref{A_aB_a}), and the second row in the table lists the number of distinct elements not from $W_0$ since $W_0$ has elements $|00\rangle$, $|01\rangle$, $|10\rangle$, $|11\rangle$. Hence, there is a QLS$(4)$ $H_{\ell}'$ in $\mathcal{H}_2^{\otimes 2}$ with $\ell$ distinct elements not from $W_0$ for $\ell\in \{2,4,6,8\}$.

Define a QLS$(8)$ of the following form:

\centerline {$\begin{pmatrix}
		|0\rangle \otimes W_0 & |1\rangle \otimes W_0\\
		|1\rangle \otimes Y & |0\rangle \otimes X\\
	\end{pmatrix},
	$}
\noindent where $X, Y$ are QLS$(4)$s having exactly $\ell_1$ and $\ell_2$ distinct elements not from $W_0$ for $\ell_1,\ell_2\in \{2,4,6,8,12$, $14,15,16\}$. Then we can obtain a QLS$(8)$ with cardinality $c\in [49,64]\setminus \{57\}$. For $c=57$, the following QLS$(8)$ has cardinality 57.  

\centerline {$\begin{pmatrix}
		|0\rangle \otimes W_0 & |1\rangle \otimes W_2\\
		|1\rangle \otimes W_4 & |0\rangle \otimes W_1\\
	\end{pmatrix}
	$}
This completes the proof. 
\qed

Similar to the proof of Lemma \ref{QLS(8)}, we present the proof of our main result.

\noindent {\bf Proof of Theorem \ref{MainResult}}:
For $m=2$, the conclusion holds by Lemma \ref{QLS(8)}. For $m\geq 3$, let $A=(a_{i,j})$, where $a_{i,j}=j-i\pmod m$. Then $A$ is a classical Latin square over $\{0,1,\ldots,m-1\}$. 
Replacing each $a_{i,j}$ with $|a_{i,j}\rangle \otimes X_{i,j}$ where $X_{i,j}$ is a QLS$(4)$ in $\mathcal{H}_2^{\otimes 2}$, we obtain a QLS$(4m)$ in $\mathcal{H}_m\otimes \mathcal{H}_2^{\otimes 2}$.
\begin{center}
	$\begin{pmatrix}
		|0\rangle\otimes X_{0,0} & |1\rangle\otimes X_{0,1}&|2\rangle\otimes X_{0,2}& \cdots &|m-1\rangle\otimes X_{0,m-1} \\
		|m-1\rangle\otimes X_{1,0} & |0\rangle\otimes X_{1,1}&|1\rangle\otimes X_{1,2}& \cdots &|m-2\rangle\otimes X_{1,m-1}\\
		|m-2\rangle\otimes X_{2,0} & |m-1\rangle\otimes X_{2,1}&|0\rangle\otimes X_{2,2}& \cdots &|m-3\rangle\otimes X_{2,m-1}\\
		\vdots & \vdots& \vdots& \ddots & \vdots
		\\|1\rangle\otimes X_{m-1,0}&|2\rangle\otimes X_{m-1,1}&|3\rangle\otimes X_{m-1,2}& \cdots &|0\rangle\otimes X_{m-1,m-1}
	\end{pmatrix}$
\end{center}
In order that this QLS$(4m)$ has cardinality $c$, we have to choose appropriate QLS$(4)$ $X_{i,j}$. 

Let $H_0,H_1,\ldots,H_8$ be defined as in the proof Lemma \ref{distinctQLS(4)}. By comparing the coordinate vectors of elements in $W_{a,b}$ in Equation (\ref{Wab}) and $H_{\ell}$ for $\ell\in \{0,1,\ldots,8\}$, it is easy to check that there are $m-1$ QSL$(4)$, say $W_{2i+3,2i+4}$ ($1\leq i\leq m-1$), in Lemma \ref{infiniteQSL(4)Cardinality16} in $\mathcal{H}_2^{\otimes 2}$ with cardinality 16 and these $16m-16$ distinct elements differ from those in $H_{\ell}$ for any $\ell\in \{0,1,\ldots,8\}$.

For $0\leq j\leq m-1$, take
$$X_{i,i+j\pmod m}\in \left \{ \begin{array}{ll}
\{H_0,H_1\} & \text{if}\ i=0\\
 \{X_{0,j},H_2,H_3,\ldots,H_8,W_{5,6}\} & \text{if}\ i=1\\
\{X_{0,j},W_{2i+3,2i+4}\} & \text{if}\ i\geq 2.\\
\end{array} \right . $$
Then all $|a_{i,i+j}\rangle \otimes X_{i,i+j}$, $0\leq i\leq m-1$, contribute to $4+4x_{j,0}+x_{j,1}+16x_{j,2}+\ldots+16x_{j,m-1}$ distinct elements where $x_{j,0}\in \{0,1\}$, $x_{j,1}\in \{0,2,3,4,5,6,7,8,16\}$ and $x_{j,i}\in \{0,1\}$ for $2\leq i\leq m-1$. Hence, this QLS$(4m)$ has totally
$$4m+4\sum_{j=0}^{m-1}x_{j,0}+\sum_{j=0}^{m-1}x_{j,1}+16\sum_{j=0}^{m-1}\sum_{i=2}^{m-1}x_{j,i}$$
distinct elements. Note that $\sum_{j=0}^{m-1}x_{j,0}$ runs through $[0,m]$, $\sum_{j=0}^{m-1}x_{j,1}$ runs through $[0,16m-8]\setminus \{1,16m-15\}$ and $\sum_{j=0}^{m-1}\sum_{i=2}^{m-1}x_{j,i}$ runs through $[0,m(m-2)]$. Hence, for any integer $c\in [4m,16m^2-8m-8]\setminus \{4m+1,16m^2-8m-15\}$, there is a QLS$(4m)$ with cardinality $c$. 

Let $W_0,W_1,W_2,W_3,W_4,H_{2}',H_4',H_6'$ and $H_8'$ be defined as in the proof of Lemma \ref{QLS(8)}. By comparing the coordinate vectors of elements in $W_{a,b}$ in Equation (\ref{Wab}) and $W_i$ for $i\in \{0,1,2,3,4\}$, $H_{\ell}'$ for $\ell\in \{2,4,6,8\}$, it is easy to check that there are $m-1$ QSL$(4)$s, say $W_{2i+3,2i+4}$ ($1\leq i\leq m-1$), in Lemma \ref{infiniteQSL(4)Cardinality16}, in $\mathcal{H}_2^{\otimes 2}$ with cardinality 16 and these $16m-16$ distinct elements differ from those in $W_i$ for $i\in \{0,1,2,3,4\}$ and those in $H_{\ell}'$ for any $\ell\in \{2,4,6,8\}$.

For $0\leq j\leq m-1$, take
$$X_{i,i+j\pmod m}\in \left \{ \begin{array}{ll}
	\{W_0\} & \text{if}\ i=0\\
	\{W_0,H_2',H_4',H_6',H_8',W_2,W_{3},W_{1},W_{5,6}\} & \text{if}\ i=1\\
	\{W_{2i+3,2i+4}\} & \text{if}\ i\geq 2.\\
\end{array} \right . $$
Then all $|a_{i,i+j}\rangle \otimes X_{i,i+j}$, $0\leq i\leq m-1$, contribute to $16+x_{j,1}+16(m-2)$ distinct elements where $x_{j,1}\in \{0,2,4,6,8,12,14,15,16\}$. Hence, this QLS$(4m)$ has totally
$$16m(m-1)+\sum_{j=0}^{m-1}x_{j,1}$$
distinct elements. Note that $\sum_{j=0}^{m-1}x_{j,1}$ runs through $[0,16m]\setminus \{1,3,5,7,9,11,13\}$. Hence, for any integer $c\in [16m^2-16m,16m^2]\setminus \{16m^2-16m+t\colon t\in \{1,3,5,7,9,11,13\}\}$, there is a QLS$(4m)$ with cardinality $c$. 

Since $16m^2-16m<16m^2-8m-8$ for $m\geq 3$, simple computation shows that the set $([4m,16m^2-8m-8]\setminus \{4m+1,16m^2-8m-15\})\cup ([16m^2-16m,16m^2]\setminus \{16m^2-16m+t\colon t\in \{1,3,5,7,9,11,13,25\}\})$ equal to $[4m,16m^2]\setminus \{4m+1\}$ if $m\neq 3$, or  $[4m,16m^2]\setminus \{4m+1,16m^2-8m-15\}$ if $m=3$. 

The following QLS$(12)$ has cardinality 105.

\centerline {$\begin{pmatrix}
		|0\rangle \otimes W_0 & |1\rangle \otimes W_0 & |2\rangle \otimes H_0\\
		|2\rangle \otimes H_6 & |0\rangle \otimes W_1 & |1\rangle \otimes W_0\\
		|1\rangle \otimes W_{5,6} & |2\rangle \otimes W_{5,6} & |0\rangle \otimes W_{5,6}
	\end{pmatrix}
	$}

This completes the proof. \qed

\section{Conclusion}

In this paper, we used sub-QLS$(4)$ to prove that for any integer $m\geq 2$ and any integer $c\in [4m,16m^2]\setminus \{4m+1\}$, there is a QLS$(4m)$ with cardinality $c$. We have tried using this method to construct other QLS$(mn)$ with possible cardinality. However, it is hard to construct QLS$(n)$s similar to our QLS$(4)$s. It is worth finding other methods to construct QLS$(n)$ with possible cardinality $c\in [n,n^2]\setminus \{n+1\}$.

\end{document}